\documentclass[reqno]{amsart}
\usepackage{amssymb,graphicx,enumitem} 
\usepackage[utf8]{inputenc}

\graphicspath{{./plots}}

\newcommand*{\eps}{\varepsilon}
\newcommand*{\bg}{\boldsymbol{g}}
\newcommand*{\bv}{\boldsymbol{v}}

\newcommand*{\e}{\mathrm{e}}
\newcommand*{\ri}{\mathrm{i}}
\renewcommand*{\d}{\mathrm{d}}
\newcommand*{\D}{\mathrm{D}}

\newcommand{\sech}{\operatorname{sech}}
\newcommand{\MSD}{\operatorname{MSD}}
\renewcommand{\Im}{\operatorname{Im}}
\renewcommand{\Re}{\operatorname{Re}}
\newcommand{\slow}{{\text{slow}}}
\newcommand{\fast}{{\text{fast}}}

\begin{document}
\title[Surrogate models for slowly driven fast oscillators]%
{Deterministic and stochastic surrogate models for a slowly driven
fast oscillator}
\date{\today}

\author[M. Oliver]{Marcel Oliver}
\author[M. Tiofack Kenfack]{Marc Tiofack Kenfack}
\address[M. Oliver and M. Tiofack Kenfack]{Mathematical Institute for Machine
Learning and Data Science \\
KU Eichst\"att--Ingolstadt \\
85049 Ingolstadt \\ Germany}

\address[M. Oliver]{Constructor University, 28759 Bremen, Germany}


\begin{abstract}
It has long been known that the excitation of fast motion in certain two-scale dynamical systems is linked to the singularity structure in complex time of the slow variables.  We demonstrate, in the context of a fast harmonic oscillator forced by one component of the Lorenz 1963 model, that this principle can be used to construct time-discrete surrogate models by numerically extracting approximate locations and residues of complex poles via Adaptive Antoulas--Andersen (AAA) rational interpolation and feeding this information into the known
``connection formula'' to compute the resulting fast amplitude. Despite small but nonnegligible local errors, the surrogate model maintains excellent accuracy over very long times.  In addition, we observe that the long-time behavior of fast energy offers a continuous-time analog of Gottwald and Melbourne's 2004 ``$0$--$1$ test for chaos'' -- the asymptotic growth rate of the energy in the oscillator can discern whether or not the forcing function is chaotic.
\end{abstract}

\maketitle

\section{Introduction}

Two-scale dynamical systems often possess adiabatic invariants, such
as the energy in the fast subsystem, which are preserved over long
periods of time \cite{Neishtadt:2008:AveragingMA}.  Typical examples
are the spring pendulum with a stiff spring, or balance in geophysical
fluid flow \cite{Van13,McI15}.  Here, we aim to model and quantify
the residual transfer of energy in and out of the fast subsystem using
information on only the slow components of the dynamics.

The slow subsystem can often be constructed as a divergent asymptotic
series in the scale separation parameter \cite{WarBokShe95} but it is
rarely an exact invariant of the full system
\cite{GelfreichL:2002:AlmostIE, GelfreichL:2003:LongPO, MacKay04}.
Optimal truncation of the asymptotic series, however, typically yields
exponentially small exchange of energy over exponentially long times
\cite{CotterR:2006:SemigeostrophicPM, Neishtadt84, Nekhoroshev77,
TemamW:2007:ExponentialAP}.  Careful asymptotic analysis in the
context of low-dimensional models reveals that the exchange of energy
between slow and fast subsystems can be linked to the presence of
complex-time singularities in the slow components and explicit
expressions for the transfer amplitude can be derived
\cite{CoxR:2003:InitializationQG, Vanneste:2004:InertiaGW, Van08,
Vanneste:04}.  The interpretation of the transfer process as a Stokes
phenomenon \cite{Berry:1988:StokesPS,OldeDaalhuisCK:1995:StokesPM}
classifies it as a time-discrete phenomenon in an asymptotic sense.

We take a new look at this old problem from the following angle: (i)
What can we say about the statistical behavior of the fast compartment
when the slow motion is chaotic, and (ii) can we use this knowledge to
construct a surrogate model that operates purely on the slow time
scale?  We answer these questions in one of the simplest possible
proof-of-concept settings: a single fast harmonic oscillator driven by
one of the components of Lorenz's 1963 model.  We do not consider any
feedback from the oscillator onto the Lorenz dynamics but rather
consider the Lorenz system as a generator for a quasi-random sequence
of complex-time singularities.  This is the simplest configuration in
which we can study these questions without being encumbered by the
details of the model as would be the case for models of intermediate
complexity or in infinite dimensions.

Our first observation is that the dynamics of the fast amplitude is
extremely well approximated by a random walk that approaches Brownian
motion in a long-time scaling when the Lorenz dynamics is chaotic and
appears quasi-periodic when the Lorenz dynamics is in a periodic
regime, a dichotomy which is the continuous-time analog of the
``$0$--$1$ test for chaos'' of Gottwald and Melbourne
\cite{GottwaldM:2004:NewTC,GottwaldM:2016:01T}.  A related dichotomy
in a continuous-time dynamical system was observed by Alvarez-Socorro
\emph{et al.}\ \cite{AlvarezCF:2019:FrontDD} in the context of
studying a pinning-depinning transition in front propagation.

Our second finding is that the full dynamics can be replicated
extremely accurately, path-wise, by a surrogate model where the
necessary information on the transition amplitudes is extracted from
the slow motion by applying rational interpolation to the numerical
trajectory and using the resulting approximate complex-time poles as
input to the ``connection formula'' which describes the jump in fast
amplitude across a Stokes line.  Therefore, the surrogate model takes
the form of a discrete walk in the complex plane for the complex
amplitude of the fast motion.  As such, the evolution of the modulus
of the fast amplitude depends on the relative phase of each additional
step, which is essentially, but not entirely, uncorrelated between
subsequent steps.  Thus, stochastic surrogate models can be
constructed by randomizing the choice of phase.

Our main tool is the Adaptive Antoulas--Anderson (``AAA'') algorithm
for rational interpolation, introduced by Nakatsukasa \emph{et al.}\
\cite{NakatsukasaST:2018:AAAAR}, which provides us with a robust
estimator for the location of the complex-time poles and their
residues of the slow dynamics.  The AAA algorithm has been applied to
a variety of different problems \cite{NakatsukasaST:2023:FirstFY},
including, recently, the study of Stokes phenomena similar to what is
done here \cite{DengL:2023:ExponentialAW,
Lustri2023ExponentialAU}.

This paper is structured as follows.  In Section~\ref{s.slow}, we
review the asymptotic theory for high-order fast-slow splitting of the
forced oscillator.  This can be done in many ways; our presentation is
in a form that easily generalizes to more complicated situations.
Section~\ref{s.matched} is a review of the derivation of the
connection formula via matched asymptotics.  Again, this material is
classical, but we provide details for the convenience of the reader
and, in particular, point out that the dynamics of the fast amplitude
can be written as a discrete walk in the complex plane.
Section~\ref{s.01} introduces the full coupled system.  By direct
numerical simulation, we observe that the solution in the fast phase
space performs approximately a discrete random walk.  We link this
observation to the $0$--$1$ test for chaos.  Section~\ref{s.aaa} gives
an overview of rational interpolation and, specifically, the AAA
algorithm.  We then test the accuracy of the AAA approximation in
conjunction with the asymptotic connection formula.  In
Section~\ref{s.deterministic}, we introduce the surrogate model,
replacing the fast components of the coupled system with the
AAA-estimator coupled with the connection formula.
Section~\ref{s.stochastic} looks at phase randomization for the
complex transition amplitude to account for the inherent uncertainty
in knowing precise phase information.  The paper concludes with a
brief discussion and outlook.

\section{The slow expansion}
\label{s.slow}

We consider a fast harmonic oscillator with a forcing function $g(t)$
that varies slowly relative to the period of oscillation,
\begin{equation}
  \eps^2 \, \ddot q(t) + q(t) = g(t) \,,
  \label{e.forced-osc}
\end{equation}
with $\eps$ a small parameter.  The solution can be written as the
linear combination of the general solution to the homogeneous
equation, which is spanned by $\exp{(\pm \ri t/\eps)}$, and a
particular solution to the inhomogeneous problem.  Let us assume that
the particular solution can be written as an asymptotic series in
$\eps$.  To leading order, the position simply follows the forcing,
i.e., $q_0(t) = g(t)$.  Higher-order terms are easily constructed, for
example by repeated integration-by-parts of the variations-of-constant
formula (see, e.g., \cite{Van08}) or, as sketched below, by direct
expansion.  The latter approach is equivalent but generalizes more
easily to systems and nonlinear equations.

Let us write $p = \eps \, \dot{q}$,
\begin{equation}
  \bv = \begin{pmatrix}
          q \\ p
        \end{pmatrix} , \qquad
  \bg = \begin{pmatrix}
          0 \\ g 
        \end{pmatrix} , \qquad \text{and } 
  J = \begin{pmatrix}
        0 & -1 \\ 1 & 0
      \end{pmatrix} .
\end{equation}
Then \eqref{e.forced-osc} takes the form
\begin{equation}
  \eps \, \dot \bv + J \bv = \bg \,.
\end{equation}
Seeking a slow solution, a solution that is slaved to $\bg$ to order
$\eps^N$, 
\begin{equation}
  \bv_\slow^N = \sum_{n=0}^{N} \eps^n \, \bv_n(\bg)
  \label{e.slow-expansion}
\end{equation}
and writing $\bv_{\fast}^N = \bv - \bv_\slow^N$, we find, by direct
computation, that
\begin{equation}
  \dot{\bv}_{\fast}^N
  = - \frac{1}{\eps}J\bv_{\fast}^N
    + \frac{1}{\eps} \, \bigl[ \bg - J \bv_0(\bg) \bigr]
    - \sum_{n=0}^{N-1} \eps^n \, \bigl[J \bv_{n+1}(\bg)
      + \D \bv_n(\bg) \dot{\bg} \bigr] + \mathcal{O}(\eps^N) \,,
  \label{e.vfast-dot}
\end{equation}
where $\D$ denotes the total derivative.  As the first term is skew,
it does not change the fast energy
\begin{equation}
  E_{\fast}^N = \lVert \bv_{\fast}^N \rVert^2 \,.
  \label{e.fast-energy}
\end{equation}
Overall, the fast energy remains invariant to $\mathcal{O}(\eps^N)$
when the intermediate terms in \eqref{e.vfast-dot} drop out, which is
the case when
\begin{equation}
  \bv_n = -J^{n+1} \, \frac{\d^n\bg}{\d t^n}
  \label{e.bvn}
\end{equation}
for $n = 0, \dots, N$.  Thus, unless $g$ is entire, the series
\eqref{e.slow-expansion} does not converge as $N \to \infty$.  Indeed,
when $g$ is analytic on the strip $\lvert \Im t \rvert < r$, we have
the Cauchy estimate
\begin{equation}
  \lvert g^{(n)} (t) \rvert
  \leq M \, \dfrac{n!}{r^n}
  \lesssim \sqrt{n} \, \Bigl( \frac{n}{r \e} \Bigr)^n
  \quad \text{with} \quad
  M = \sup_{\lvert \Im z \rvert < r} \lvert g(z) \rvert \,,
\end{equation}
the second bound due to Stirling's approximation.  When $g$ has poles,
this bound is essentially sharp, so that an optimal truncation of the
series is achieved by choosing $N \sim r/\eps$ (see, e.g.,
\cite{Dingle:1973:AsymptoticE}), and
\begin{equation}
  \eps^N \, \lvert g^{(N)} (t) \rvert
  \lesssim \frac{1}{\sqrt \eps} \, {\e}^{-\tfrac{r}\eps} \,.
  \label{e.last-term-bound}
\end{equation}
As we shall compute in the next section, and already noted in
\cite{Van08}, the remainder of the series is larger, by a factor of
$\sqrt\eps$, than bound \eqref{e.last-term-bound} on the last included
term, but is still exponentially small.  We also see that the
influence of the remainder on the evolution of the fast energy is
strongest when $t$ is closest to a complex-time singularity of $g$,
which is the time when the bounds above saturate.  A careful analysis
is given in the next section.

\section{Matched asymptotics}
\label{s.matched}

Let us first assume that $g(t)$ has a single pair of simple poles, i.e.
\begin{equation}
  g(t) = \frac{a}{t-t_\star} + \text{entire} + \text{c.c.}
  \label{e.gsimplepole}
\end{equation}
where the entire remainder gives rise to a converging slow expansion,
thus does not contribute to the excitation of fast motion, and there
is a complex conjugate contribution since $g$ is assumed to be
real-valued.  To begin, we neglect all but the first term on the right
of \eqref{e.gsimplepole}.

Away from the pole at $t=t_\star$, the solution to
\eqref{e.forced-osc} can be approximated by a superposition of the general solution to the harmonic oscillator and the leading order of the slow expansion, i.e.,
\begin{equation}
  q(t) = A \, \e^{\ri t/\eps} + B \, \e^{-\ri t/\eps}
         + \frac{a}{t-t_\star} + O(\eps)
\label{outer_sol1}
\end{equation}
for constants $A$ and $B$ to be determined later.

Near $t=t_\star$, this outer solution breaks down.  A solution in the inner region is obtained by introducing scaled variables
\begin{equation}
  \tau = \frac{t - t_\star}{\eps}
  \quad \text{and} \quad
  \phi(\tau) = \frac{\eps \, q(t)}{a} \,,
\end{equation}  
so that \eqref{e.forced-osc} reads
\begin{equation}
  \phi'' +\phi = \frac{1}{\tau} \,,
\label{eq_inner}
\end{equation}
where the superscript prime denotes differentiation with respect to
$\tau$.  A formal solution is given by
\begin{equation}
  \phi(\tau)
  = \frac{1}{2\ri} \,
    \bigl[
      \e^{-\ri\tau} \, E_1(-\ri\tau) - \e^{\ri\tau} \, E_1(\ri\tau)
    \bigr]
  \label{e.phi}
\end{equation}
where
\begin{equation}
  E_1 (z) = \int_z^\infty \frac{\e^{-1}}t \, \d t
\end{equation}
is known as the \emph{exponential integral}, as can be verified by
direct computation.  This solution is valid in any region of the complex plane where the right-hand side of \eqref{e.phi} is analytic.
Since $E_1$ has a logarithmic singularity at the origin, we need to
make an appropriate choice of the associated branch cut.  Assuming,
without loss of generality, that $\Im t_\star>0$, we wish to obtain a
solution that is valid in the half plane $\{ \Im \tau < 0 \}$.
Placing the branch cut of $E_1(z)$ onto the negative imaginary axis
and working in the right half plane $\Re z > 0$, we observe that the
argument of the second exponential integral in \eqref{e.phi} is always
in the required half-plane.  When $\tau = x - 0\ri$ with $x<0$, the
argument of the first exponential integral is away from the branch
cut, so can be moved into the right half plane by continuity.  As
$E_1(z) \sim \e^{-z}/z$ for $|\arg(z)|<\frac{3\pi}{2}$ as
$\lvert z \rvert \to \infty$ \cite[equation
5.1.51]{AbramowitzS:1972:HandbookMF}, this implies
$E_1(\ri\tau) \sim \e^{-\ri\tau}/(\ri\tau)$ and
$E_1(-\ri\tau)\sim \e^{\ri\tau}/(-\ri\tau)$, so that
\begin{equation}
  \phi(\tau) \sim \frac{1}{\tau}
  \quad \text{for } \tau < 0 \,.
  \label{eq_inner_sol_negativ}
\end{equation}
When $\tau = x - 0\ri$ with $x<0$, the argument of the first
exponential integral, $z=- \ri \tau$, crosses over into the left half
plane along the branch cut, and we need to use the identity
\begin{equation}
  E_1(-\ri (\tau - 0 \ri)) = E_1(-\ri \tau + 0) + 2 \pi \ri \,.
\end{equation}
Then $E_1(-\ri\tau) \sim \e^{\ri\tau}/(-\ri\tau) + 2\pi\ri$ and
$E_1(\ri\tau) \sim \e^{-\ri\tau}/(\ri\tau)$, so that
\begin{equation}
  \phi(\tau) = \frac{1}{\tau} + \pi \, \e^{-\ri\tau}
  \quad \text{for } \tau > 0 \,.
  \label{eq_inner_sol_possitiv}
\end{equation}
Matching the inner expansions \eqref{eq_inner_sol_negativ} and
\eqref{eq_inner_sol_possitiv} with the outer expansion
\eqref{outer_sol1} in an intermediate asymptotic regime gives $A=B=0$
for $\Re t \ll 0$, resp.\ $A=0$ (consistent with the expectation that there are no exponentially large terms) and
\begin{equation}
  B = \frac{\pi \, a}{\eps} \, \e^{\ri t_\star/\eps} 
  \label{eq_amplitude_fast_oscillations}
\end{equation}
for $\Re t \gg 0$.  Taking into account the contribution from the
complex conjugate pole at $\Bar{t}_\star$, we see that, as $t$ crosses
the ``Stokes line'' between the two poles, the oscillatory
contribution
\begin{equation}
  q_{\fast}
  \sim \frac{2\pi}{\eps} \Re \bigl[
         a \, \e^{\ri(t_\star - t)/\eps}
       \bigr]
  \label{eq_fast_part}
\end{equation}
as $\eps \to 0$ is ``switched on'', here again with the convention
that $\Im t_\star>0$ so that the fast contribution is exponentially
small in $\eps$.

This expression agrees with that of Vanneste \cite[equation
3.6]{Van08}, who performed a careful analytic and numerical study on a
single such emission event and found that the predicted asymptotic
connection amplitude is in very good agreement with numerics, even for
moderately small values of $\eps$.  He also demonstrated that the
growth of fast amplitude in the inner region follows the profile of an
error function.  Here, however, we are not interested in the local
details of a single emission event, but rather look for the cumulative
effect of many events.  Seen from the long time scale, each event
contributes a discrete jump to the fast energy.

Let us thus assume that $g(t)$ has $m$ pairs of complex conjugate
simple poles with residues $a_i$ and locations $t_i$ with the
convention that $\Im t_i>0$, for $i=1, \dots, m$, as well as their
complex conjugate residues and locations.  Identifying the amplitude
of the vector $\bv_{\fast} = (q_{\fast}, p_\fast)$ with a point in the
complex plane, \eqref{eq_fast_part} shows that each pair of complex
conjugate poles contribute a complex connection amplitude
\begin{equation}
  B_i = \frac{2\pi}{\eps} \, a_i \, \e^{\ri t_i/\eps} \,.
  \label{e.connection-formula}
\end{equation}
The overall complex amplitude after crossing $m$ Stokes lines is given
by linear superposition, i.e.,
\begin{equation}
  B(m) = \sum_{i=1}^m B_i \,.
  \label{e.many-poles}
\end{equation}
Thus, the dynamics of the fast amplitude can be described as a
discrete walk in the complex plane, each step exponentially small in
$\eps$ and linked to the complex-time poles of $g$ via
\eqref{e.connection-formula}.  In the following, we are interested in
the long-time asymptotics as $m \to \infty$.

\begin{figure}
\centering
\includegraphics[width=\linewidth]{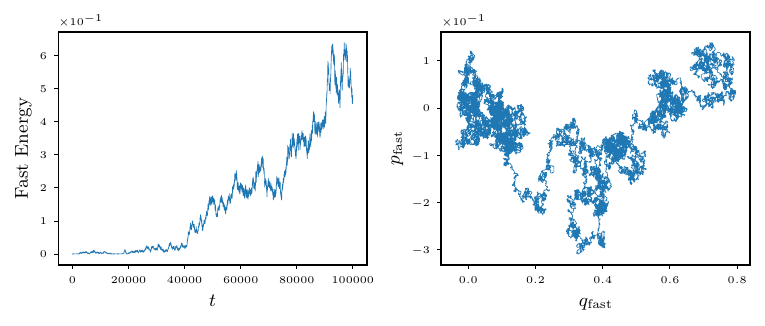}
\caption{Fast oscillator driven by the Lorenz system in the chaotic
regime with standard parameters $b=\tfrac83$, $r=28$, and $\sigma=10$.
We take $\eps = 0.01$ and diagnose to order $N=16$.}
\label{fig_fast_M_chaotic}
\end{figure}

\begin{figure}
\centering
\includegraphics[width=\linewidth]{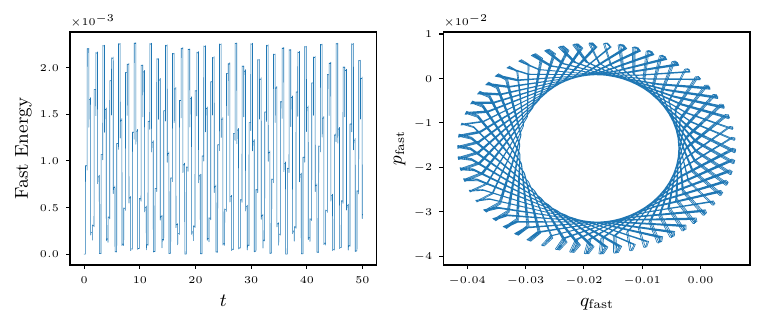}
\caption{Fast oscillator driven by the Lorenz system in the periodic
regime, here point A1 of \cite{DullinSR:2007:ExtendedPD} where
$b=\tfrac83$, $r=402$, and $\sigma=302$.  We take $\eps=0.001$
and diagnose to order $N=11$.}
\label{fig_fast_M_periodic}
\end{figure}

\section{Periodic vs.\ chaotic forcing}
\label{s.01}

We employ the Lorenz 1963 model \cite{Lorenz1963} as a driver for the
forced oscillator equation \eqref{e.forced-osc}.  Originally derived
by truncating the equations for Rayleigh--B\'{e}nard convection to
lowest harmonics, the system is now a standard test case for concepts
involving low-dimensional chaos.  It reads
\begin{subequations}
\begin{align}
  \dot{x} &= \sigma(y-x) \,, \\
  \dot{y} &= r x -y -xz \,, \\
  \dot{z} &= xy - \tfrac{8}{3}z \,.
\end{align}
\label{eq_lorenz_system}
\end{subequations}
In our context, the singularity structure of solutions on the
attractor is crucial.  Foias \emph{et al.}\
\cite{FoiasJK:2001:LorenzEM} have shown that the attractor has an
uniform radius of real analyticity, i.e., the singularities remain
bounded away from the real time axis.  A formal power count shows that
if there are poles, they should be expected to be simple in $x$ and of
second order in $y$ and $z$.  Viswanath and Şahutoğlu
\cite{ViswanathS:2010:ComplexSL} show that there are indeed solutions
of this type, albeit with corrections given by logarithmic psi-series
(``quasi-poles'').  It remains open whether all singularities are
locally of this type.  Webb \cite{Webb:2013:ComputingCS} has shown
that the singularity structure of solutions to the Lorenz equations
can be approximated via rational interpolation.  Moreover, he found
that the singularities are located close to the time at which the
$z$-component has a local maximum.  Thus, even though the Lorenz
system is not known to strictly fit into the framework laid out in
Section~\ref{s.matched}, its singularity structure appears to be close
enough to make it an interesting test case.

We choose to drive the oscillator with the $z$-component of the Lorenz
system, i.e.\ $g(t) = z(t)$.  We first initialize the Lorenz system
arbitrarily and run forward in time until the orbit is practically on
the attractor.  Then we initialize the oscillator in an optimally slow
state via expression \eqref{e.slow-expansion}.  Its coefficients
\eqref{e.bvn} are computed via repeated analytic differentiation of
the Lorenz equations, a task that is easily performed by computer
algebra.  The optimal order truncation $N$ was determined empirically.
The coupled system is then evolved forward in time.

As diagnostics, we plot the fast energy as given by
\eqref{e.fast-energy}.  It shows clear jumps which, as we shall see
later, are associated with complex-time singularities of $z(t)$.  We
also plot a Poincar\'e section of the orbit of $\bv_\fast^N$ which is
obtained by sampling $\bv_\fast^N(t)$ at integer multiples of the fast
period. For the standard values of the Lorenz parameters where the
motion on the attractor is chaotic, the samples of the phase point are
performing a pseudo-random walk in the fast phase space, see
Fig.~\ref{fig_fast_M_chaotic} where it is evident that the fast
energy, on average, is growing in time.

When the parameters for the Lorenz system are changed so that the
attractor is a single periodic orbit, the fast phase point follows a
quasi-periodic motion and the fast energy remains bounded
(Fig.~\ref{fig_fast_M_periodic}).

This dichotomy is reminiscent of the ``$0$--$1$ test for chaos'' of
Gottwald and Melbourne \cite{GottwaldM:2004:NewTC,GottwaldM:2016:01T}.
The test asks whether a given, for simplicity infinite, scalar time
series $g_n$ is ``regular'' or ``chaotic''.  The time series is used
to drive the discrete-time dynamical system
\begin{subequations}
  \label{e.01}
\begin{align}
  p_{n+1} & = p_n + g_n \, \cos(cn) \,, \\
  q_{n+1} & = q_n + g_n \, \sin(cn) \,,
\end{align}
\end{subequations}
for some fixed frequency parameter $c$. From there, one computes the
\emph{mean square displacement}
\begin{equation}
  \MSD(n) = \lim_{N\to \infty}  \sum_{j=1}^N
    \frac{\lvert p_{j+n} - p_j \rvert^2 +
      \lvert q_{j+n} - q_j \rvert^2}N \,.
\end{equation}
Under mild assumptions, the \emph{asymptotic growth rate} exists and
generically takes one of two values:
\begin{equation}
  K = \lim_{n \to \infty} \frac{\log \MSD(n)}{\log n}
  =
  \begin{cases}
  0 & \text{iff } g_n \text{ regular} \,, \\
  1 & \text{iff } g_n \text{ chaotic} \,.
  \end{cases}
\end{equation}
In the exceptional case when $g_n$ is periodic and the frequency $c$
is resonant, secular growth, corresponding to a quadratic increase of
the mean square displacement is possible.  A rigorous justification
of this behavior is given in \cite{GottwaldM:2009:Validity01}.

Here, we note that the $0$--$1$ test is actually a discrete version of
the driven harmonic oscillator: Writing \eqref{e.forced-osc} in
co-rotating coordinates and discretizing with the explicit Euler
scheme of unit step size gives \eqref{e.01}.

\section{Numerical detection of poles: The AAA algorithm}
\label{s.aaa}

To link the behavior seen in the previous section with the singularity
structure of the forcing function $g$, we need to continue the
numerical solution, which is known only for a discrete set of real
$t$, into the complex plane.  The problem of analytic continuation is
ill posed (see, e.g., \cite{Trefethen:2020:QuantifyingIC}), but can be
stabilized by implicit or explicit regularization.  One such algorithm
is the Adaptive Antoulas--Anderson (AAA) method, first introduced by
Nakatsukasa \emph{et al.}\ \cite{NakatsukasaST:2018:AAAAR} and found
to perform extremely well for the problem of analytic continuation
\cite{trefethen2023numerical}.  Even though a complete theoretical
analysis remains open, the observed excellent performance makes it the
algorithm of choice for our purposes.

The starting point for the AAA algorithm is the barycentric rational
interpolation formula \cite{Salzer:1981:RationalII,
SchneiderW:1986:NewAR} for function values $f_1, \dots, f_m$ given on
support points $z_1, \dots, z_m$,
\begin{equation}
  r(z) = \frac{n(z)}{d(z)}
       = \sum_{k=1}^{m} \frac{w_k \, f(z_k)}{z - z_k}
         \bigg/ \sum_{k=1}^{m} \frac{w_k}{z - z_k} \,.
  \label{eq_aaa.interpolante}
\end{equation}
The weights $w_1, \dots, w_m$ are arbitrary at this point, but
non-zero.  Continuously extending the function at the support points,
we see that $r(z_k) = f_k$ for $k = 1, \dots, m$.  The poles of $r$
lie elsewhere; they can be computed efficiently by solving a
generalized eigenvalue problem
\cite{klein2012applications,NakatsukasaST:2018:AAAAR}.  The
barycentric interpolation formula has excellent numerical stability.
However, the global behavior of the interpolant, in particular the
singular set, depend greatly on the choice of the weights and may be
very unstable as a function of the data.

There are various strategies for choosing the weights
\cite{Berrut:1988:RationalFG, BerrutT:2004:BarycentricLI,
FloaterH:2007:BarycentricRI}.  In particular, Antoulas and Anderson
\cite{antoulas1986scalar} proposed to consider the rational
\emph{approximation} problem for sample points
$\Gamma = \{z_1, \dots, z_M\}$ with sample values $f_1, \dots, f_M$
and select a subset $\Gamma_m \subset \Gamma$ with $m\ll M$ as support
points on which the barycentric formula interpolates, then optimize
the weights by solving the least-squares problem
\begin{equation}
  \text{minimize } \lVert fd - n \rVert_{\Gamma\setminus\Gamma_m}
  \quad \text{subject to } \lVert w \rVert_m = 1 \,,
  \label{e.optimization}
\end{equation}
where $\lVert \, \cdot \, \rVert_{\Gamma\setminus\Gamma_m}$ is the
discrete $2$-norm on the set of sample points which are not
interpolation support points and $\lVert \, \cdot \, \rVert_m$ is the
discrete $2$-norm on $m$-vectors.  This problem can be solved
efficiently by singular value decomposition.  However, the question
remains how large $m$ should be and how to choose the set $\Gamma_m$.

AAA uses a greedy selection procedure: start with a minimal set of
support points $\Gamma_m$, solve the optimization problem, and check
if the overall approximation error is acceptable.  If it is not, pick
the sample point with the largest pointwise error, add it to
$\Gamma_m$, and repeat.  This procedure typically converges very fast,
e.g.\ root-exponentially even for functions with branch points
\cite{TrefethenNW:2021:ExponentialNC}.

\begin{figure}
\centering
\includegraphics{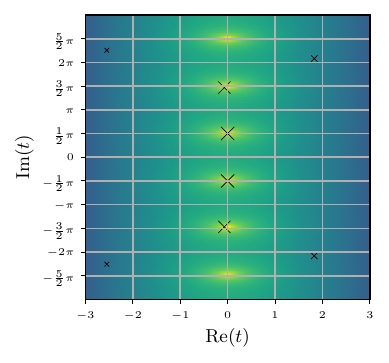}
\caption{AAA-estimate of the location of the poles of
$\operatorname{sech}(t)$.}
\label{fig_aaa_sech_estimate}
\end{figure}

\begin{figure}
\centering
\includegraphics[width=\textwidth]{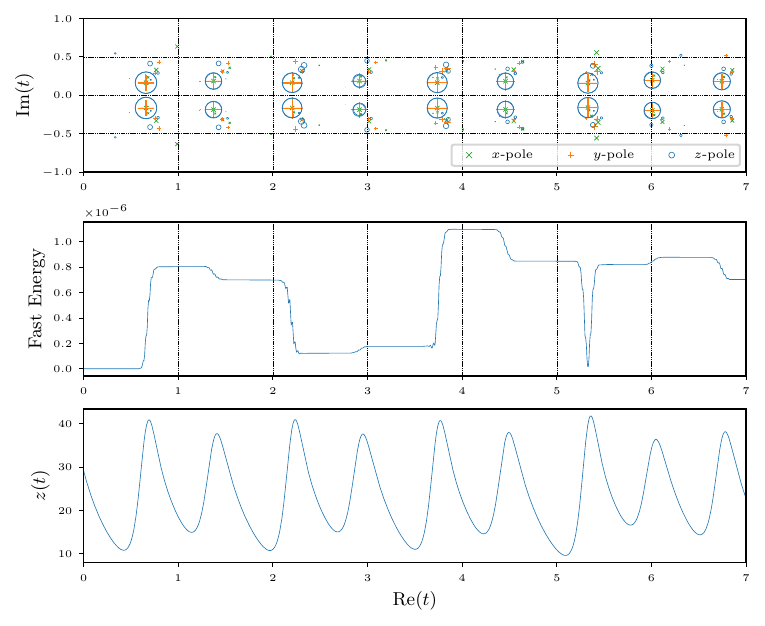}
\caption{Approximation of the singularities of the Lorenz system via
AAA rational approximation with $M=1000$ sample points in the interval
$[0,7]$ (top), the corresponding evolution of fast energy of the
forced oscillator with $\eps = 0.01$ (middle), and the $z$-component
of the Lorenz system driving the oscillator (bottom).}
\label{fig_matching_pole_jumps}
\end{figure}

Fig.~\ref{fig_aaa_sech_estimate} shows an approximation to the
location of the poles of the function $g(t) = \sech(t)$, which is the
standard example in \cite{Van08}, using the AAA algorithm on $50$
equidistant real sample points in the interval $[-15,15]$ with the
restriction $m \leq m_{\text{max}}=9$.  The colored background
indicates the modulus of the true function, crosses indicate the
estimated locations of the poles.  The pair of poles closest to the
real axis is reproduced with good accuracy, the location of the second
pair is already slightly inaccurate, and further out, the estimates
are completely off.  The figure also shows that AAA does not impose
any symmetries on the results, even when the data has an axis of
symmetry.

For our purposes, this behavior is perfectly acceptable, as it is only
the poles close to the real axis that matter.  To put some numbers on
this statement: on the $g(t) = \sech(t)$ test case with
$\eps = \tfrac14$, certainly not a very small number, the asymptotic
formula \eqref{e.connection-formula} applied with the exact values for
the two innermost pairs of poles coincides with an estimate by direct
numerical simulation, where the fast energy is diagnosed to order
$N=15$, up to a relative error of $0.00003\,\%$.  Using only the
innermost pair of poles, accuracy drops to $0.0004\,\%$.  Replacing
exact pole locations and residues with a AAA estimate based on $50$
sample points, accuracy is ``only'' $0.001\,\%$; in this setting,
using more than one pair of AAA poles will not help.  Increasing the
number of sample points given to the AAA algorithm from $50$ to $100$,
accuracy improves to $0.0002\,\%$, apparently compensating some of the
error of the connection formula.  In this case, even though in this
case the contribution from the second pole is fairly accurate, it does
not help to improve the error, but neither does it hurt to include
these contributions or even include the contributions from all further
AAA poles.

One concern when using AAA is the possible emergence of spurious poles
with a residue of order machine precision or poles so close together
that they almost cancel each other out, which are called Froissart
doublets \cite{NakatsukasaST:2018:AAAAR}.  These poles do not matter
here as they do not contribute substantially to the connection
amplitude \eqref{e.connection-formula} when the contributions of all
poles are summed up.  However, as we shall see later, when breaking up
the forcing function into smaller pieces, the AAA approximation may
produce poles that are close to the real axis, but outside of the
sampling interval.  These poles must be removed, for else they will
contribute large spurious connection amplitudes.

For the Lorenz model, the singularity structure is conjectured to
consist of complex conjugate pairs of branch points
\cite{ViswanathS:2010:ComplexSL, Webb:2013:ComputingCS}.  The AAA
approximation, on the other hand, will always construct a sequence of
simple poles, as can be seen in Fig.~\ref{fig_matching_pole_jumps}
(upper panel).  Here, the discrete pole with the largest residue is
located close to the presumed branch point, with further poles of
smaller residue further out in the complex plane, indicating the
possible location of a branch cut.  Close inspection of the data shows
that each of the dominant poles in the $y$- and $z$-components close
to the real axis is actually a pair of twin poles with almost the same
location and residue.  This is consistent with the fact that these
singularities are formally expected to be of order two, cf.\ the
discussion in Section~\ref{s.01}.  The fact that the AAA algorithm
works so well for branch points is analyzed in detail in a recent
preprint by Lustri \emph{et al.}\ \cite{Lustri2023ExponentialAU}.
Here, we observe similarly that for singularities which are close to
poles of order two, the effective connection amplitude can be
decomposed into a sum of connection amplitudes of the form
\eqref{eq_amplitude_fast_oscillations}, each associated with a simple
pole, with high accuracy.

In this preliminary test, we have used the AAA algorithm on the entire
time interval without restricting the number of interpolation nodes.
The approximation is also done separately on the different components
of the model, as in \cite{Webb:2013:ComputingCS}, which leads to
different locations of the poles, especially those further out in the
complex plane, even though the location of singularities must be the
same in all three components.  In principle, this could be enforced by
performing the optimization problem \eqref{e.optimization}
simultaneously on the entire solution vector.  In
Section~\ref{s.deterministic}, however, we go a different route: we
will use AAA approximation only on the $z$-component, cut the solution
into small chunks, and put a hard limit on the number of poles.

Fig.~\ref{fig_matching_pole_jumps} also shows the agreement with the
horizontal location of the poles and the jumps in the fast energy of
the forced oscillator (middle) and the correlation with the maxima of
$z$ (bottom).

We recall that, according to \eqref{e.many-poles}, the complex
amplitude of the generated oscillation performs a discrete
pseudo-random walk in the complex plane, with a step when a new Stokes
line is crossed.  Thus, the modulus of the fast amplitude increases
maximally when the phase of the new connection amplitude is aligned
with the current phase of the fast amplitude, it decreases maximally
when the phases are anti-aligned but can do anything in between
depending on the difference in phase.  This is clearly visible in
Fig.~\ref{fig_matching_pole_jumps} where the big jumps require a pole
relatively close to the real axis, but not every pole that is close to
the real axis contributes a big jump. This sort of superposition is
also discussed in the context of exponential asymptotics for nonlinear
lattice waves in \cite{DengL:2023:ExponentialAW}.

In a chaotic regime, pseudo-randomness of the continuous-time
trajectories implies pseudo-randomness of the (discrete) sequence of
singularities.  Thus, we can use the AAA approximation of a Lorenz
trajectory as a generator of either a pseudo-random or quasi-periodic
sequence of complex conjugate pairs of poles.

\section{Deterministic surrogate model}
\label{s.deterministic}

We now seek to replace the $5$-dimensional dynamical system consisting
of the Lorenz equations and the driven oscillator by a simpler one
where only the slow dynamics are maintained as an ODE and the
evolution of the fast amplitude is computed via the cumulative
connection amplitude \eqref{e.many-poles}, where the terms are
computed from the approximate discrete set of singularities of the
$z$-component of the Lorenz system.  The surrogate system is simpler
not only because continuous-time dynamics is only three-dimensional
but also, in particular, because it evolves only on the slow time
scale.  To solve the coupled $5$-dimensional system, off-the-shelf
time-stepping schemes would require time steps on the order of the
fast period.  This is prohibitive when $\eps$ is small.  Of course,
for this simple problem, there are various exponential time-stepping
strategies that could maintain a time step uniform in $\eps$.
However, these need to be problem-adapted and do not generalize well.

Since we know that the singularities of the Lorenz system occur near
the maxima of $z$ and nowhere else, we can construct the surrogate
model as follows.
\begin{enumerate}[label={\upshape(\roman*)}]
\item Take a (slow) numerical solution of the Lorenz model,
\item \label{i.2} cut the discrete time series of the $z$-variable
into pieces contained in intervals
$[t_k^{\text{start}}, t_k^{\text{end}}]$ conditioned on the bound
$z(t) \geq z_\star$,
\item feed the time series from each piece into the AAA algorithm
limited to order $m \leq m_{\max}$ to obtain a list of poles and
residues for each interval,
\item remove spurious poles, defined as poles whose real part lies
outside of the interpolation interval
$[t_k^{\text{start}}, t_k^{\text{end}}]$,
\item compute the local complex connection amplitude
\begin{equation}
  B_k = \frac{2\pi}\eps \sum_{(a,z) \in \mathcal P}
          a \, \e^{\ri z/\eps} \,,
  \label{e.local}
\end{equation}
where $\mathcal P$ denotes the set of tuples $(a,z)$ that characterize
the remaining set of complex-conjugate pairs of poles, with $a$
denoting the residue and $z$ the location of the member of the pair
that has $\Im z>0$.
\end{enumerate}
The global evolution of the fast amplitude is finally given by the
cumulative connection formula \eqref{e.many-poles}.

\begin{figure}
\centering
\includegraphics[width=0.8\textwidth]{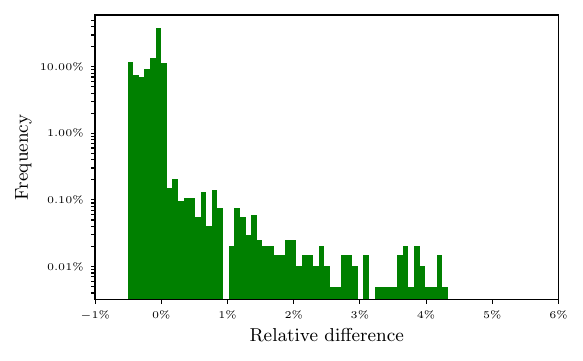}
\caption{Distribution of the difference between the fast amplitude from
direct numerical evaluation and the fast amplitude from the
connection formula at time $t_k^{\text{end}}$, relative to the largest
such amplitude, when the fast component is re-initialized to a
balanced state at every $t_k^{\text{start}}$.  The mean of this
distribution is at a relative difference of $-0.15\,\%$.}
\label{fig_connection_formula}
\end{figure}

To check whether the approximation of the change in fast amplitude via
the local connection formula \eqref{e.local} is accurate, we first
perform a pre-test in which we compare each $B_k$ with a direct
numerical solution as follows: We initialize the coupled model in a
balanced state at each $t_k^{\text{start}}$ via
\eqref{e.slow-expansion}, evolve forward in time, and diagnose the
fast amplitude as the square root of the fast energy
\eqref{e.fast-energy} at $t_k^{\text{end}}$.
Fig.~\ref{fig_connection_formula} shows the resulting distribution of
the relative error, where positive values indicate that the fast
amplitude from direct numerical simulation is larger than $B_k$, and
vice versa.  The distribution is highly peaked within a band of about
$0.5\,\%$, but has a longer tail toward positive values.  Thus, errors
are small but far from negligible, with a non-zero mean.

We note that the error is not uniform as $\eps \to 0$.  The reason is
that the AAA algorithm replaces the second-order branch point by a
collection of poles that have close, but not identical real parts of
their locations.  Since these real parts determine the phase of the
associated connection amplitude, they are coherent only so long as the
period of the fast oscillation is long compared to the phase jitter
introduced by the approximation.  As $\eps \to 0$, coherence will
eventually be lost and the approximation degrades.  This effect has
been studied and explained in detail in a recent preprint
\cite{Lustri2023ExponentialAU}.  Increasing the maximal number of
poles will lead to better behavior for small values of $\eps$.  We
conjecture that uniform-in-$\eps$ behavior can be obtained in our
setting by modifying the AAA algorithm to force all poles to have the
same real part.

\begin{figure}
\centering
\includegraphics[width=\linewidth]{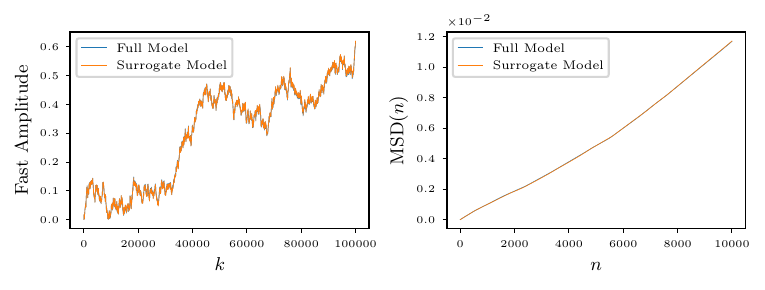}
\caption{Comparison between the full model with the surrogate model in
the chaotic regime for $\eps = 0.01$. The two trajectories are
visually almost identical over a time period covering $100\,000$
return events.}
\label{fig_comparison_long_realization_chaos}
\end{figure}

We then let both models evolve freely.
Fig.~\ref{fig_comparison_long_realization_chaos} compares the direct
numerical simulation to the surrogate model, showing remarkably good
agreement over a large number of events.  To explain how this is
possible, we first note that the coupled model and the surrogate model
use the identical numerical trajectory for the Lorenz subsystem.
Thus, the rapid exponential divergence of Lorenz trajectories is not
at play here.  However, if we were using even slightly different
numerical settings for the two cases, we would see entirely different
trajectories on the time scale shown. 

Perhaps more surprising is the fact that when both models evolve
freely, the agreement is much better than the amplitude error
statistics shown in Fig.~\ref{fig_connection_formula} would suggest.
One reason is the following: When $B(k)$ is nonzero for some $k$, the
effect of adding the next complex jump amplitude $B_{k+1}$ to the
current fast amplitude $B(k)$ depends on the relative phase between
the two.  Due to the scale separation, this relative phase is
essentially random and uniformly distributed, so that the effect of
the error in $B_{k+1}$ on the new amplitude $\lvert B(k+1) \rvert$ is
distributed nearly symmetrically about $0$ when
$\lvert B(k) \rvert \gg \lvert B_{k+1} \rvert$.  Thus, the error in
$B(k)$ can be considered as a random walk on top of the pseudo-random
walk $B(k)$, thus grows like $\sqrt k$ in expectation, not like $k$ as
could be expected naively.  However, even this explanation does not
fully account for the remarkable long-time accuracy of the surrogate
model.  Numerically, we observe a strong remnant temporal correlation
of the surrogate model error across neighboring jump events. The
precise mechanism remains an open question.

As a key statistical measure we compute, inspired by the ``$0$--$1$
test'' described in Section~\ref{s.01}, the finite-time mean square
displacement
\begin{equation}
  \MSD(n) = \sum_{j=1}^N
    \frac{\lvert p_{j+n} - p_j \rvert^2 +
      \lvert q_{j+n} - q_j \rvert^2}N \,,
\end{equation}
with $p_j= \eps \, \dot q_{\text{fast}}(t_j^{\text{end}})$ and
$q_j= q_{\text{fast}}(t_j^{\text{end}})$, where $N$ is the number of
$z$-excursions as separated out in Step~\ref{i.2} above.  As suggested
in \cite{GottwaldM:2016:01T}, we restrict to $n \leq N/10$; for larger
values of $n$, the finite-time mean-square displacement will deviate
significantly from its time-asymptotic value.

\begin{figure}
\centering
\includegraphics[width=\linewidth]{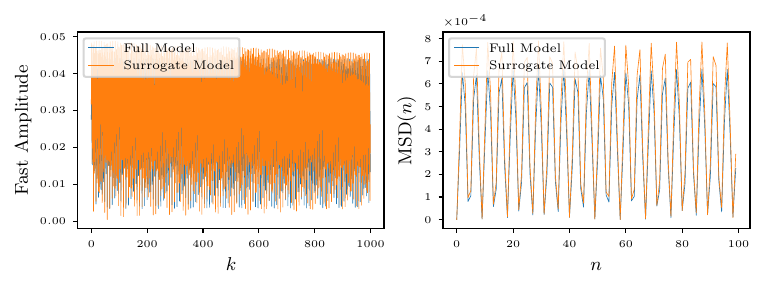}
\caption{Comparison between the full model with the surrogate model in
the periodic regime with $\eps = 0.001$.}
\label{fig_realization_periodic1}
\end{figure}

In a parameter regime where the Lorenz dynamics is periodic, the
resulting fast amplitude is quasi-periodic, both for the full system
and for the surrogate system (Fig.~\ref{fig_realization_periodic1}).
Moreover, the MSD remains bounded, as expected from the ``$0$--$1$
test'', with only minor quantitative differences between the two
models.

\section{Stochastic surrogate model}
\label{s.stochastic}

Even though we have seen that the reconstruction of the full complex
connection amplitude, including phase information, is remarkably
successful, it may be too much to ask for from a modeling
perspective.  An $O(\eps)$ perturbation to the state of the Lorenz
system will cause an $O(1)$ change in phase for the connection
amplitude already from one Stokes line event to the next.  Thus,
regarding the ability of the model to predict the future, the phase
may be considered effectively random.

In the present setting, we can easily modify the connection formula by
adding an independently uniformly distributed random phase to each
term in the cumulative connection formula, defining
\begin{equation}
  B_{\text{rand}}(m) = \sum_{i=1}^m B_i \, \e^{\ri\alpha_k}
  \quad \text{with } \alpha \sim \mathcal{U}_{[0,2\pi]} \,.
\end{equation}
Within the formula for the local connection amplitude \eqref{e.local},
the phase must be kept coherent, as the set of local poles is merely
an approximation to a single pair of branch points.

\begin{figure}
\centering
\includegraphics[width=\linewidth]{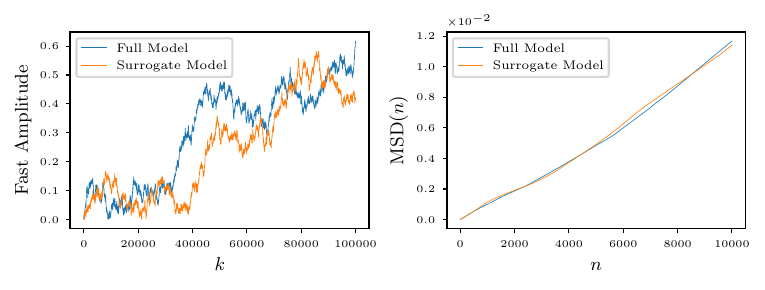}
\caption{Comparison between the full model with the surrogate model with
phase randomization in the chaotic regime, with $\eps = 0.01$.}
\label{fig_realization_chaotic2}
\end{figure}

\begin{figure}
\centering
\includegraphics[width=\linewidth]{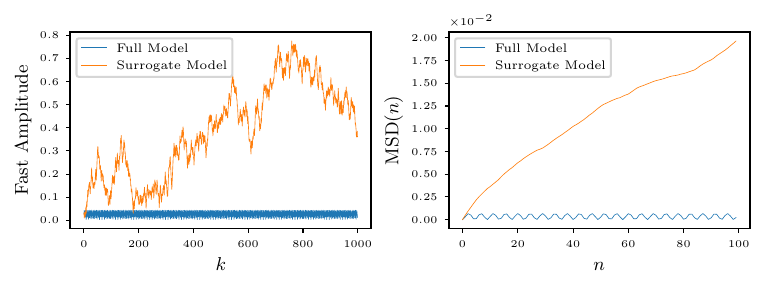}
\caption{Comparison between the full model with the surrogate model with
phase randomization in the periodic regime, with $\eps = 0.001$.}
\label{fig_realization_periodic2}
\end{figure}

This stochastic surrogate model, of course, does not follow the
development of the fast amplitude path-wise, but also has a linear
growth of mean-square displacement in the chaotic regime, with
approximately the same slope (Fig.~\ref{fig_realization_chaotic2}).
When the phase of the connection amplitude is randomized with periodic
driving, the behavior of fast energy is statistically that of the fast
energy in the chaotic driving regime
(Fig.~\ref{fig_realization_periodic2}).

\section{Discussion}

We have shown that the discrete structure of singularities in complex
time of the forcing function suffices to describe the level of energy
in a forced fast harmonic oscillator, locally with good accuracy and
excellent long-time behavior due to the fact that local errors
partially cancel over a long time scale.  Thus, a discrete surrogate
model that takes the singularity structure of the forcing function as
input suffices to represent the evolution of fast energy with high
accuracy.

The distance on the real axis between poles, linked to the relative
phase of the contribution to the fast energy of each pole dictates
whether or not an upcoming pole will contribute constructively or
destructively.  While this phase is uniformly distributed, for finite
scale separation even a chaotic forcing signal has enough remnant
phase correlations that they are visible in the long-time statistics
of the evolution of the fast energy.  Thus, complete randomization of
the phase will underestimate the mean square displacement of the fast
amplitude (or energy).

We further observed that the long-time behavior of the fast energy of
a forced harmonic oscillator can be used to detect whether the forcing
signal is chaotic or not, in complete analogy with the ``$0$--$1$ test
for chaos'' \cite{GottwaldM:2004:NewTC,GottwaldM:2016:01T}.  More
generally, this work gives an example of how even exponentially small
phenomena can accrue significant contributions over a large number of
events, which means that they may not be negligible and must be
modeled.

So far, we have explored one of the simplest possible settings.  We
believe that it will be easy to extend these ideas to related
low-dimensional systems where a connection formula is known, such as
those considered in \cite{Vanneste:2004:InertiaGW, Van08,
Vanneste:04}.  Alternative approaches for low-dimensional models based
on machine learning have, e.g., been developed in
\cite{ChekrounLM:2021:StochasticRF}.

More challenging is the extension to the geophysical balance problem
where the setting is infinite-dimensional and where it is not even
clear whether global linear wave and vortical modes are the right
basis in which to describe the problem.  In addition, there may be
nonlinear fast-fast interactions which imply that the simple linear
superposition of the contribution of each pole does not strictly hold
true (see the discussion in \cite{Lustri2023ExponentialAU}).
Nonetheless, we conjecture that the linear growth of the mean square
displacement of the amplitude of the fast degree of freedom seen here
generalizes to the linear growth of mean square displacement of
inertial gravity wave amplitudes in geophysical flows forced by
turbulent eddy fields.  This type of ``spontaneous generation'' of
waves may therefore provide a non-negligible contribution to the
energy cycle worth further study.

\section*{Acknowledgments}

We thank Mickaël Chekroun, Sergey Danilov, Georg Gottwald, Stephan
Juricke, Edgar Knobloch, Anton Kutsenko, Nick Trefethen, Jacques
Vanneste, and Nedjeljka Žagar for stimulating discussions and remarks
on various aspects of this work.  We further thank the anonymous
referees for insightful comments and the suggestion of further highly
relevant references. Our numerical experiments use the implementation
of the AAA algorithm by Clemens Hofreither
\cite{Hofreither:2021:AlgorithmBR, Hofreither:2022:BarycentricRA}
which we gratefully acknowledge.  This paper is a contribution to
project L2 of the Collaborative Research Center TRR 181 ``Energy
Transfers in Atmosphere and Ocean'' funded by the Deutsche
Forschungsgemeinschaft (DFG, German Research Foundation) under project
number 274762653.

\bibliographystyle{siam}
\bibliography{trr.bib}

\end{document}